\documentclass[prb,aps,twocolumn,showpacs,amsmath,amssymb]{revtex4-1}
\usepackage{float}
\usepackage{amsmath,bm}
\usepackage{dsfont}
\usepackage[dvips,final]{graphicx}
\usepackage{epsfig}
\usepackage{xcolor}
\usepackage[normalem]{ulem}

\begin{document}

\title{Composition-driven magnetic anisotropy and spin polarization in Mn$_2$Ru$_{1-x}$Ga Heusler alloy}

\author{Ram\'on Cuadrado}
\address{Instituto de Ciencia de Materiales de Madrid, CSIC, Cantoblanco 28049 Madrid, Spain}

\date{\today}

\begin{abstract}
We present a comprehensive investigation of the influence of Ru concentration on the lattice parameters, atomic magnetic moments, electronic structure, and magnetic anisotropy energy of the full Heusler L2$_1$-type Mn$_2$Ru$_{1-x_p}$Ga alloy, where $x_p = 0.0834 \cdot p$ with $p = 0, \dots, 12$. This study combines first-principles calculations with data-driven techniques from artificial intelligence, specifically principal component analysis~(PCA), to reveal trends and correlations across multiple structural, magnetic, and electronic descriptors. For each composition, a set of inequivalent atomic configurations was fully optimized. Structurally, the relaxed lattices exhibit anisotropic expansion, with a pronounced elongation of the out-of-plane lattice parameter ($c$) relative to the in-plane lattice vectors, which promotes the development of perpendicular magnetic anisotropy. In addition, our analysis demonstrates that the spatial arrangement of Ru vacancies, forming correlated vacancy pairs and clusters, plays a key role in stabilizing the out-of-plane magnetization by breaking local lattice symmetry and enhancing spin–orbit coupling~(SOC). As a result, an out-of-plane easy axis emerges at intermediate Ru concentrations~(25–58\%), whereas low and high Ru levels favor an in-plane orientation or even vanishing anisotropy. The half-metallic character is also modulated by Ru content, appearing selectively at both ends of the composition range. Additionally, the ferrimagnetic coupling between Mn(4a) and Mn(4c) sublattices leads to nearly compensated magnetic moments below 50\% Ru content, with a net moment close to zero around 30\%. These findings open a pathway toward the design of tunable spintronic materials with co-optimized perpendicular magnetic anisotropy and half-metallicity, making Mn$_2$RuGa a promising candidate for magnetic tunnel junctions, magnetoresistive random-access memory~(MRAM) devices, and high-density magnetic storage applications.
\end{abstract}


\maketitle

\section{Introduction}\label{intro-sec}

Full Heusler alloys with the general formula X$_2$YZ, where X and Y are typically transition metals and Z is a non-magnetic {\it sp} element, constitute a versatile class of magnetic ternary intermetallic compounds. In particular, those crystallizing in the inverse L2$_1$ structure exhibit high structural symmetry and chemical tunability, which give rise to a wide array of magnetic, electronic, and magnetotransport properties. These characteristics make them ideal candidates for next-generation functional materials~\cite{heusler4}. Among their most attractive features for spintronic applications is the potential for half-metallicity—where one spin channel behaves as a metal while the other is semiconducting—resulting in nearly 100\% spin polarization at the Fermi level. This property enables high spin-filtering efficiency in magnetic tunnel junctions, spin valves, and MRAM devices~\cite{kurt2014}. Additionally, many Heusler alloys obey the Slater–Pauling rule, allowing for predictive control of the net magnetization via valence electron count and compositional tuning~\cite{yang2015}.

The Mn$_2$RuGa system serves as a prototypical example among Heusler compounds, combining ferrimagnetic order with half-metallic behavior. Its structure includes two Mn sublattices occupying Wyckoff positions 4a and 4c, which are antiferromagnetically coupled. The imbalance between their local moments enables tunable net magnetization, with complete magnetic compensation achievable at specific Ru concentrations~\cite{betto2015}. Experimental investigations on Mn$_2$Ru$_{1-x_p}$Ga thin films deposited on MgO(100) substrates with varying Ru content ($x_p$$\approx 0.5$–$0.9$) have revealed a composition-dependent compensation temperature~(temperature where opposing sublattice magnetizations cancel, resulting in zero net magnetization.), ultrahigh coercivity near compensation, and a decrease in spin polarization at higher Ru concentrations~\cite{siewierska2021}. Furthermore, magnetic anisotropy energy~(MAE) is a key property for spintronic device performance and can arise in Heusler and related alloys through lattice tetragonality or interfacial strain. Even slight deviations from cubic symmetry may induce an out-of-plane magnetization axis, essential for thermal stability and efficient magnetization switching in MRAM and spin-transfer-torque applications~\cite{madhura2023,klemmer2014,klemmer2016,toyota2021}. While MAE tuning via strain and composition has been extensively explored in various Heusler systems, a detailed understanding of the interplay between Ru content, lattice distortion, and MAE evolution in Mn$_2$RuGa remains limited~\cite{raja2016,sicong2022}. In addition to anisotropy, Mn$_2$Ru$_{1-x_p}$Ga alloys display rich electronic behavior: increasing Ru concentration modifies the net magnetic moment (MM) in accordance with Slater–Pauling-like trends, while the half-metallic character persists at low Ru fractions and degrades at higher values~\cite{banerjee2020}. Moreover, ultrafast all-optical toggle switching has recently been demonstrated in this same system without the need for rare-earth elements, enabling reversible magnetization reversal in under 2~ps near the magnetic compensation point\cite{banerjee2020}. This breakthrough emphasizes the importance of understanding both static and dynamic magneto-electronic properties as a function of Ru concentration—highlighting how precise compositional tuning governs not only equilibrium properties like half-metallicity and MAE, but also enables ultrafast control mechanisms relevant to next-generation spintronic technologies.

Over the last decade, Heusler compounds have been extensively investigated through first-principles calculations due to their remarkable tunability of structural, electronic, and magnetic properties. High-throughput computational studies have explored large families of Heusler alloys to identify materials with optimized MAE, half-metallicity, and spin-transport properties, revealing systematic composition–property relationships across broad chemical spaces \cite{Marathe2023, Jiang2021, Sanvito2017}. In parallel, ab initio investigations have addressed the electronic structure and magnetization dynamics of Heusler systems in technologically relevant phenomena such as all-optical switching, where the interplay between band structure, exchange interactions, and SOC plays a central role~\cite{Zhang2022}. Considerable attention has also been devoted to Heusler-based heterostructures and thin films, particularly interfaces with MgO, due to their relevance for magnetic tunnel junctions and spintronic devices. These studies have analyzed interfacial electronic transport, spin polarization, and magnetic stability in various Heusler/MgO systems~\cite{Jiang2022, Bhattacharya2023, Ma2021, Roy2021}, highlighting the sensitivity of magnetic and transport properties to interface chemistry and structural relaxation.

Beyond ideal ordered compounds, several theoretical works have emphasized the role of lattice distortions and chemical disorder in determining the magnetic response of Heusler alloys. In particular, tetragonal distortions have been shown to strongly influence MAE and electronic structure \cite{Buchelnikov2019, Faleev2017}, while correlation effects and atomic site disorder can significantly modify MMs and spin polarization. For Mn-based Heusler alloys, previous ab initio studies have demonstrated that deviations from ideal lattice ordering, especially involving Mn site occupations, can strongly impact the resulting ferrimagnetic behavior and magnetic compensation.~\cite{Galanakis2014} Within this broader theoretical framework, the present work focuses on the compositional tuning of Mn$_2$Ru$_{1-x_p}$Ga, combining first-principles calculations with data-driven analysis to systematically correlate lattice distortion, electronic structure, and MAE across a wide range of Ru concentrations.

The main objective of the present work is to investigate how varying the Ru content in the inverse full Heusler alloy Mn$_2$RuGa affects its MAE, lattice parameters, and electronic structure. To this end, we construct a comprehensive ensemble of inequivalent atomic configurations spanning the full substitution range and carry out full geometry and electronic structure optimization using density functional theory (DFT). This methodology allows us to elucidate how Ru depletion leads to anisotropic lattice distortions (i.e., changes in $c/a$), drives a reorientation of the magnetic easy axis from in-plane to out-of-plane, alters sublattice MMs, and modulates the degree of half-metallicity. Moreover, our computational predictions regarding the evolution of MAE and spin-polarized electronic structure across the Mn$_2$Ru$_{1-x_p}$Ga series are in excellent agreement with experimental findings reported by Fowley {\it et al.}\cite{fowley2018}. That study showed that these films retain high spin polarization for Ru concentrations below $x_p \leq 0.7$, with magnetic compensation and MAE emerging in specific composition–strain regimes. Additionally, anomalous Hall effect measurements on nearly compensated films confirmed a large coercive field and high anisotropy energy density (approximately 216~kJ/m$^3$), consistent with a strong out-of-plane easy axis in the compensated state. 

Finally, to uncover hidden patterns and correlations among the structural, electronic, and magnetic properties beyond direct inspection, we employ principal component analysis (PCA), implemented in Python (scikit-learn). This dimensionality-reduction approach enables us to identify the most relevant combinations of descriptors and to reveal the dominant trends that govern the evolution of magnetic properties across the Mn$_2$Ru$_{1-x_p}$Ga series. In general, the use of unsupervised machine-learning techniques, including PCA and related dimensionality-reduction methods, has recently gained considerable attention in condensed-matter physics as a powerful tool to detect phase transitions, classify complex phases, and identify hidden correlations in large datasets. Previous works have successfully applied these approaches to both classical and quantum systems, including the identification of phase transitions~\cite{Wang2016,Jadrich2018,Marashli2025} and crossover regimes~\cite{Hu2017}, as well as the analysis of nonequilibrium dynamical behavior~\cite{McDermott2020}, vortex matter systems~\cite{Reichhardt2025} and quantum phases~\cite{Lidiak2020}. Within this broader context, the present work employs PCA as a physically interpretable framework to correlate structural distortions due to Ru vacancies, electronic structure, and MAE across compositional space. In this sense, PCA provides a bridge between first-principles calculations and descriptor-based analysis, enabling the identification of physically meaningful correlations that can guide the rational optimization of magnetic properties.

The paper is organized as follows: Section~\ref{theory-sec} outlines the construction of atomic configurations, the filtering protocol for identifying inequivalent structures, and the computational methodology. It also includes the subsection~\ref{scope-sec}, where the scope and robustness of the computational approach are analyzed. Section~\ref{geom-sec} presents the results of structural optimization based on DFT calculations. Section~\ref{mm-sec} analyzes the evolution of MMs across the composition range. Section~\ref{dos-sec} discusses the projected density of states (PDOS), while Section~\ref{mae-sec} focuses on the influence of chemical composition on MAE. Section~\ref{pca-sec} examines the trends revealed by the PCA and elucidates how they relate to the underlying physical properties of the system. Finally, Section~\ref{conclusions-sec} summarizes our main findings. 

\section{Theoretical Methods}\label{theory-sec}

\begin{figure}
\centering
\epsfig{file=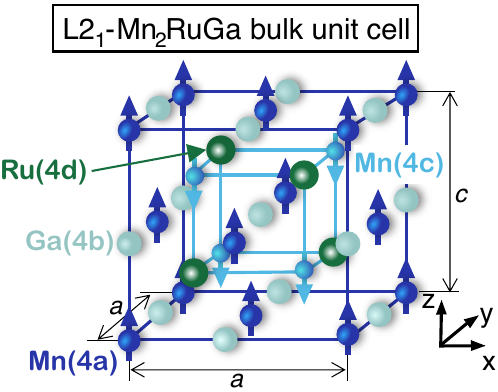,scale=0.8}
\caption{(Colour online) Schematic representation of the full Heusler L2$_1$--type Mn$_2$RuGa bulk alloy. Dark and light blue spheres represent Mn atoms occupying inequivalent crystallographic Wyckoff positions~(4a and 4c, respectively), while light cyan and green spheres denote Ga~(4b) and Ru~(4d) atoms, respectively. Wyckoff positions are indicated in brackets, and arrows superimposed on Mn atoms illustrate the orientation of their local MMs.}\label{mrg-fig}
\end{figure}

Full Heusler compounds with the general formula X$_2$YZ can crystallize in four structural types: regular cubic, regular tetragonal, inverse cubic, and inverse tetragonal~\cite{sergey}. In the present study, we focus on the inverse cubic Heusler structure. In this configuration, the X atoms occupy two crystallographically distinct Wyckoff positions—4a (0, 0, 0) and 4c (1/4, 1/4, 1/4)—while the Y and Z atoms are located at 4d (3/4, 3/4, 3/4) and 4b (1/2, 1/2, 1/2), respectively. In our case, Mn is assigned to the X sites, Ru to the Y site, and Ga to the Z site (see Fig.~\ref{mrg-fig}).

To systematically investigate the effect of Ru concentration on the geometric, electronic, and magnetic structure of the L2$_1$--type Mn$_2$RuGa alloy we conduct a comprehensive first-principles study of Mn$_2$Ru$_{1-x_p}$Ga alloy, where $x_p$=0.0834$\cdot$$p$ with $p = 0, \dots, 12$. To accommodate this range of Ru concentrations, the simulation cell was expanded threefold along the $z$-axis, resulting in a 48-atom supercell containing 12 Ru atoms at full occupancy ($x_p$=0 $\Leftrightarrow$ $p$=0), which corresponds to 100\% Ru content. This supercell size represents a compromise between computational cost and configurational sampling, enabling the description of multiple Ru concentrations while capturing diverse local atomic environments. Within the present high-throughput and statistical framework, finite-size effects are expected to be smaller than the intrinsic configurational variability included in the Boltzmann-weighted analysis. Starting from the full Mn$_2$RuGa structure, Ru atoms were progressively removed one by one to explore reduced compositions. We also assume that the local crystal structure of Mn$_2$RuGa remains unchanged, meaning that atomic substitutions between different species are not allowed, which implies that substitutional disorder is not explicitly considered. A fully random treatment would require substantially larger configurational sampling and is therefore beyond the scope of the present study. The removal of Ru atoms creates vacancies that permit various atomic rearrangements, as the remaining Ru atoms can occupy any of the 12 available lattice sites. For a fixed number of vacancies $p$, the number of possible configurations corresponds to the binomial coefficient $C(12,p)$=$12!/p!(12-p)!$, where $p=0$ represents the fully ordered structure~(no Ru atoms removed) and $p = 12$ corresponds to complete Ru depletion. The number of distinct configurations for $p = 0$ to $p = 6$ are: 1, 12, 66, 220, 495, 792, and 924, respectively; for $p = 7$ to $p = 12$, the sequence mirrors this distribution, yielding a total of 4096 possible crystal structures. To reduce computational cost, we performed single--point self-consistent field~(SCF) energy calculations for all 4096 configurations. Symmetry--equivalent structures were identified by comparing their total SCF energies with a precision threshold of 1$\cdot$10$^{-3}$~eV. This filtering process reduced the number of inequivalent configurations to 603--a significantly more tractable set for scalar-relativistic~(SR) ionic optimization and fully-relativistic~(FR) calculations.
%New for the reply

All geometrical, electronic, and magnetic quantities in this work, unless explicitly stated otherwise, correspond to statistical averages over the different phases observed at each Ru concentration, $p$, weighted by their relative stabilities. The partition function  $\mathcal{Z}_p$ is given by:
\[
\mathcal{Z}_p = \sum_{l=1}^{N_p} exp\big[\Delta E_{vf,p}^l/kT]
\]
where  $l$ indexes the total number of configurations for a given composition  $p$~(denoted as  $N_p$), and  $\Delta E_{vf,p}^l$  represents the energy difference between the highest vacancy formation energy~(VF) among configurations with the same Ru content,  $E_{vf,p}^{\max}$, and the VF of each specific configuration  $l$,  $E_{vf,p}^l$. The latter is defined as:

\[
E_{vf,p}^l = E_p^l - ( n_{Ru}\cdot E_{sc}^{Ru} + n_{Mn}\cdot E_{sc}^{Mn} + n_{Ga}\cdot E_{sc}^{Ga})
\]
where  $E_p^l$  is the total SCF energy for each $l,p$, while  $E_{sc}^{(Ru,Mn,Ga)}$ represents the total energy of Ru, Mn, and Ga in their most stable bulk phases. The terms  $n_{(Ru,Mn,Ga)}$ denote the number of atoms of each element within the unit cell for the given $p$. For any quantity  $A$, such as the projected density of states and MAE, its statistical average for each $p$ is calculated as:
\[
\langle A_p\rangle = \mathcal{Z}_p^{-1} {\textbf A^T_p} \cdot {\textbf B_p}
\]
Here,  $\textbf{A}^T_p$  and  $\textbf{B}_p$  are  row and column $N_p$--dimensional vectors representing the values  $A^l_p$  and the Boltzmann weight  $\exp\big[\Delta E_{vf,p}^l/k T\big]$, respectively. On the other hand, the fictitious Boltzmann factor assigns a higher or lower probability to each configuration depending on temperature: at higher temperatures, configurations with larger energy differences become more accessible, whereas at lower temperatures, those with lower energy differences are favored. In the present work, we assigned two distinct values to $kT$, namely 5~meV and 1~eV, in order to probe the behavior of the Mn$_2$Ru$_{1-x_p}$Ga alloy across configurations with varying degrees of metastability. The lower value~(5~meV) corresponds to $\sim$50~K and allows the analysis of the most energetically favorable configurations. In contrast, the higher value (1 eV) extends the sampling to include less stable configurations and should be interpreted as a statistical sampling parameter rather than a physically realistic thermodynamic temperature. Within this framework, the use of kT = 1 eV provides an upper-limit sampling scenario that enables evaluation of the robustness of the magnetic and electronic trends against configurational disorder. This approach provides a broader understanding of possible magnetic behaviors and mimics, in an approximate manner, the wide distribution of local atomic environments that may arise during non-equilibrium growth or synthesis processes.

Our density functional based calculations were performed using the SIESTA code~\cite{siesta1, siesta2}. Core electrons were described using fully separable Kleinman–-Bylander and norm--conserving pseudopotentials~(PPs) of the Troullier–-Martins type~\cite{kb, tm}. For the PPs we used the following optimized core radii and atomic configurations: 2.61, 0.66, 3.39 for Mn~(4s$^2$3d$^5$4p$^0$), 2.81, 1.14, 2.96 for Ru~(5s$^1$4d$^7$5p$^0$) and 1.79, 2.57, 3.00 for Ga~(4s$^1$4p$^2$4d$^0$). As exchange--correlation~(XC) potential we have employed the generalized gradient approximation~(GGA) in the Perdew, Burke, and Ernzerhof~(PBE) version~\cite{pbe} and to address a better description of the magnetic behavior, nonlinear core corrections were included in the XC terms~\cite{cc}. As a basis set, we have used a double--$\zeta$ polarized~(DZP) strictly localized numerical atomic orbitals. The electronic temperature~(kT in the Fermi-Dirac distribution) was set to 10~meV for all the calculations. To achieve a more precise description of the geometric, electronic and magnetic structure, we performed a SC optimization of the basis set and PPs cutoff radii for Mn, Ru and Ga. This was accomplished using an in--house Python--based automated workflow with SIESTA as core tool and the total SCF energy serving as the global optimization variable. We used the Mulliken partitioning scheme to obtain charge distribution \cite{mulliken}. 

The structural optimization was performed using the conjugate gradient~(CG) method at spin--polarized and SR level and until the interatomic forces were below 0.02 eV/\AA\ and the stress tolerance for the lattice vectors reached 1$\cdot$10$^{-4}$~eV/\AA$^3$. To calculate the MAE, we have performed FR calculations where the SOC is included as implemented in the SIESTA code~\cite{LSpaper, LSpaper2}. In addition to the conventional definition of MAE as the difference in total SCF energy between hard and easy magnetization directions, we computed the total energies along multiple spin orientations to systematically assess anisotropy. Due to the small contribution of SOC energy to the total energy, the level of precision required to perform an accurate FR self--consistent calculation is quite demanding. This is specially true for the calculation of MAEs, where energy differences between two spin--quantization axis are typically in the meV~(and sub--meV) range. In such calculations, the tolerance in the SC criteria~(either related to the Hamiltonian, density matrix, or both), the k--point sampling or the size of the real--space grid must be carefully converged for each specific system to ensure accurate results. Within the present work, we performed an exhaustive analysis of the convergence of these quantities until the difference between consecutive SC energy values were less than 1$\cdot$10$^{-4}$~eV within the first Brillouin zone. As a result, we set up 11$\times$11$\times$4 for the k--points sampling (see section~\ref{mae-sec} for convergence tests) and 900~Ry for the real-space grid.
%New for the reply

As part of our data-driven methodology, we employ techniques from artificial intelligence and multivariate data analysis to reveal patterns and correlations within complex datasets. PCA is one such technique, widely used to reduce the dimensionality of multivariate data while retaining the most significant sources of variation. In the present study, we have constructed a descriptor matrix $\mathbf{D} \in \mathbb{R}^{n \times m}$, where n is the number of inequivalent crystal structures and m the number of descriptors. The descriptor set was selected to capture the key structural, magnetic, and electronic factors controlling lattice distortion, magnetic compensation, and SOC-driven anisotropy. Specifically, for each crystal structure, we have considered nineteen descriptors: the difference between in-plane and out-of-plane lattice constants ($\Delta_{c-a}$), the average MM per Mn sublattice, the total MM, the Ru content, the total SCF energy differences between various spin–quantization axes: between X and Z ($\Delta E_{x-z}$), between Y and Z ($\Delta E_{y-z}$), between XY and Z ($\Delta E_{xy-z}$), and between Y and X ($\Delta E_{y-x}$). In addition, to capture the effect of vacancy distribution after Ru extraction and local geometric anisotropy, we included descriptors quantifying this spatial arrangement of Ru vacancies: $N_{\rm pairs}^{\rm tot}$ represents the total number of correlated vacancy pairs, defined as pairs of vacancies whose separation falls below a given distance threshold, thus indicating a spatially correlated arrangement within the unit cell and its periodic images. $N_{\rm pairs}^\parallel$ counts only those pairs where both vacancies lie approximately in the same crystallographic plane (in-plane pairs), while $N_{\rm pairs}^\perp$ counts pairs where vacancies are primarily separated along the out-of-plane (vertical) direction. $N_{\rm pairs}^{3D}$ includes all pairs that are separated in three dimensions, without restriction to a specific plane.
\begin{figure}
\centering
\epsfig{file=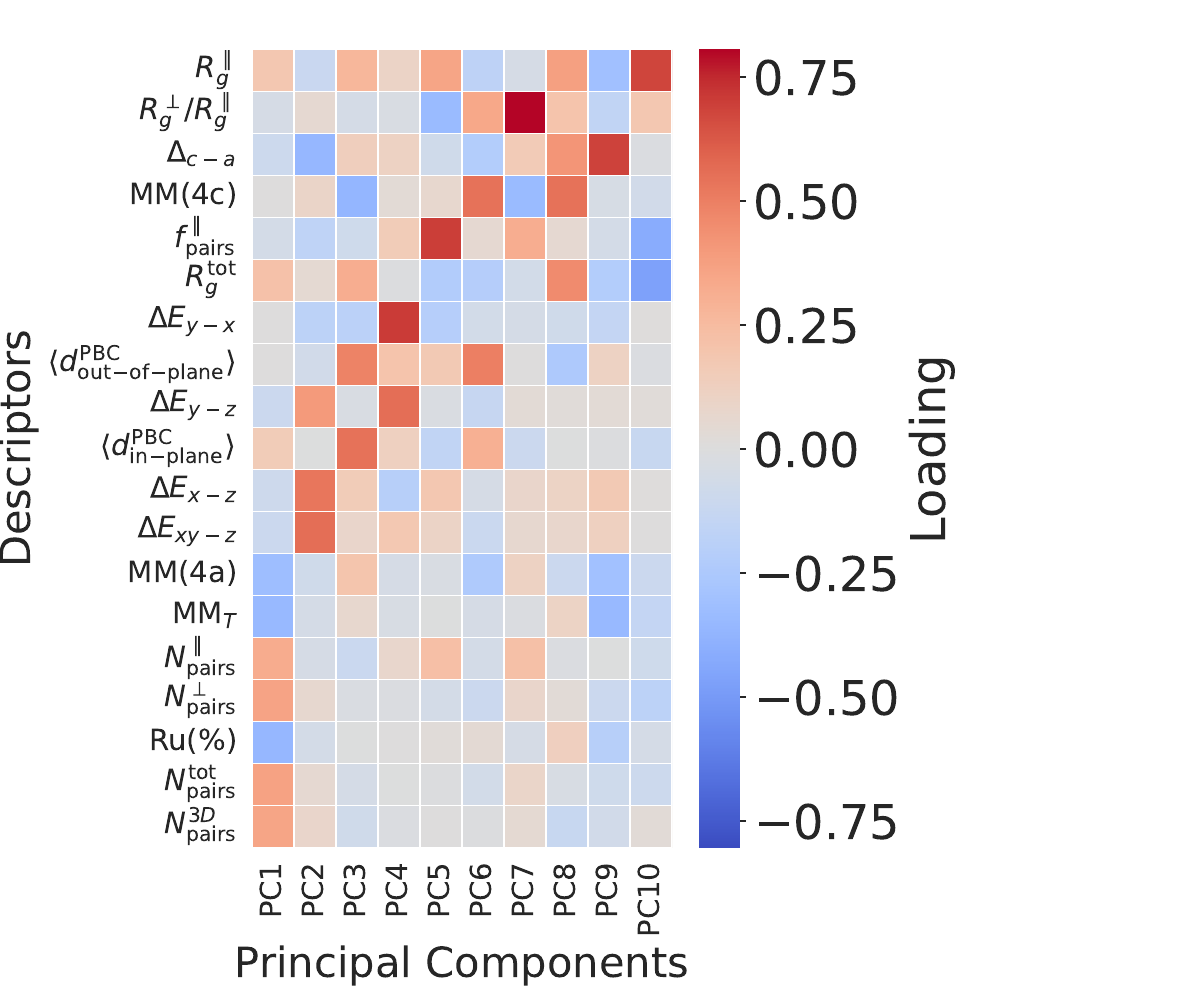,scale=0.52}
\caption{(Colour online) Heat map of the PCA loadings for the selected descriptors of Mn$_2$Ru$_{1-x_p}$Ga. Each cell represents the contribution of a descriptor to a given principal component (PC1–PC10). High absolute values indicate strong influence on the corresponding component, allowing identification of the descriptors driving the observed trends in structural and electronic properties.} \label{heat-map-fig}
\end{figure}
The fraction of in-plane pairs, $f_{\rm pairs}^\parallel = N_{\rm pairs}^\parallel / N_{\rm pairs}^{\rm tot}$, provides a normalized measure of the planar alignment of vacancies. The planar radius of gyration ($R_g^\parallel$) describing how spread out the vacancies are within the plane, the anisotropy of the gyration tensor ($R_g^\perp/R_g^\parallel$) measuring the vertical versus planar spread, the total radius of gyration ($R_g^{\rm tot}$) as a measure of overall spatial dispersion, the mean pairwise distances between vacancies along in-plane and out-of-plane directions ($\langle d^{\rm PBC}_{\rm in-plane} \rangle$, $\langle d^{\rm PBC}_{\rm out-of-plane} \rangle$), where PBC denotes periodic boundary conditions used to account for interactions between vacancies across neighboring supercells. All these descriptors capture the degree of clustering, alignment, and anisotropy of the vacancy network, which can significantly affect the local crystal-field symmetry and hence the MAE. It is worth to noting that although PCA is sensitive to the choice of descriptors, the selected variables are physically motivated and directly measurable within the DFT framework, ensuring interpretability of the resulting principal components. Since these descriptors have different units and numerical ranges, the first step is to center and standardize each descriptor. Centering ensures that each variable has zero mean, while standardization scales all descriptors to unit variance, preventing variables with larger magnitudes from disproportionately influencing the PCA results. Once the descriptor matrix $\mathbf{D}$ has been centered and standardized, PCA identifies orthogonal axes (principal components, PCs) capturing the largest variance in the data. Projecting the original data onto these axes provides a reduced-dimensional representation, revealing trends and correlations that may be obscured in the full descriptor space. To determine the number of PCs required to capture over 95\% of the total variance in the descriptors, we analyzed the scree plot (not shown here), which displays the variance explained as a function of the number of PCs and serves as a guide for component selection. Based on this analysis, we retained ten PCs, with explained variance contributions decreasing as follows: 36\%, 14\%, 11\%, 8\%, 7\%, 6\%, 5\%, 4\%, 3\% and 2\% for PC1 through PC10, respectively. It is worth to mention that although higher-order PCs explain progressively smaller fractions of the total variance, their loadings were analyzed to assess whether they capture secondary correlations between electronic structure and MAE. These components mainly describe subtle variations associated with local atomic arrangements and do not modify the dominant compositional trends identified in PC1 and PC2. An equally important outcome of PCA is the set of loadings, which represent the coefficients of the original descriptors in each PC and quantify the strength of their contribution. Since the PCs are ranked according to the variance they explain, descriptors with large loadings in the first few components dominate the primary trends in the data. These loading patterns can be visualized through a heat map, which highlights the descriptors most responsible for the observed trends (see Fig.~\ref{heat-map-fig}). 

\begin{figure*}
\centering
\epsfig{file=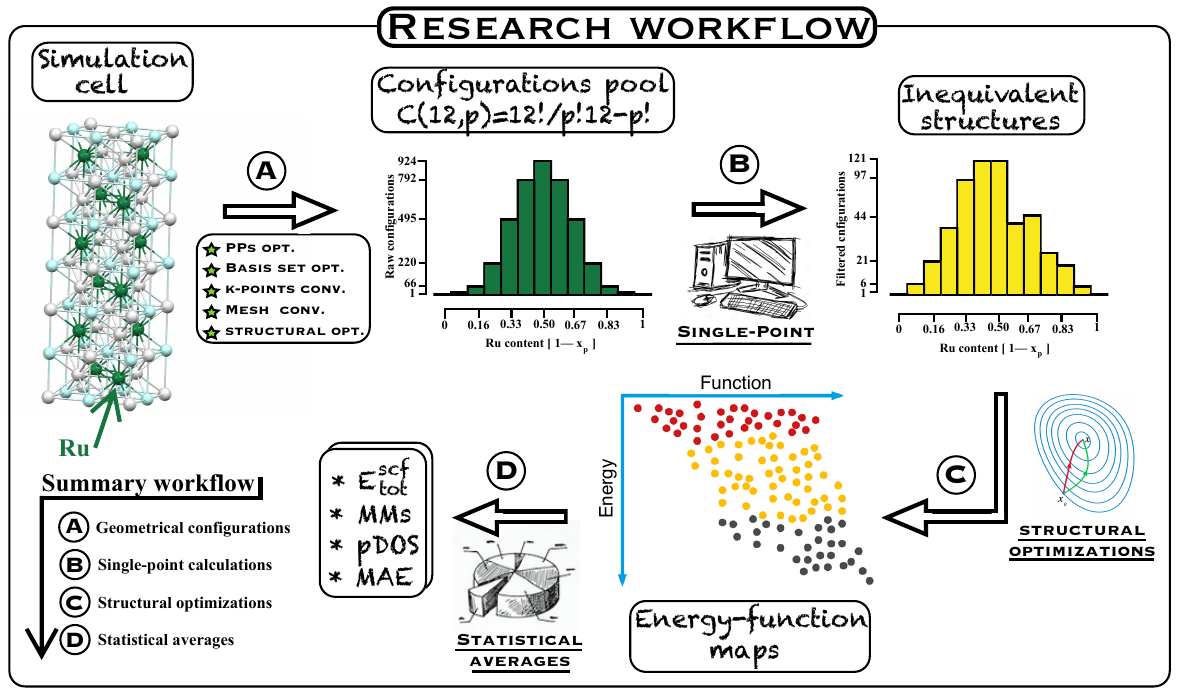,scale=0.80}
\caption{(Colour online) Schematic research workflow implemented in the present investigation. From left to right: (A) The full Heusler Mn$_2$RuGa simulation cell is first used to optimize atomic pseudopotentials and numerical orbitals, as well as to converge the $k$-point sampling and mesh cutoff; (B) Ru atoms (solid green spheres) are then selectively removed to generate thousands of possible geometric configurations for each Ru concentration. Single-point energy calculations are subsequently performed to identify and discard symmetrically equivalent structures, retaining only the inequivalent ones. The schematic yellow bars shown between steps B and C represent, in a qualitative manner, the number of inequivalent configurations obtained after this filtering procedure for each Ru composition, rather than the number of symmetry-equivalent vacancy arrangements; (C) These inequivalent configurations are fully relaxed through SR ionic optimization; (D) Once convergence is achieved, statistical averages can be computed for each Ru concentration $1-x_p$, or all configurations can be directly analyzed to extract trends in the targeted physical properties.} \label{wf-fig}
\end{figure*}

%New for the reply

\subsection{Scope and robustness of the computational approach}\label{scope-sec}

The present study employs the SIESTA code, which uses a linear combination of atomic orbitals (LCAO) and norm-conserving PPs to perform DFT calculations. Because the reliability of any DFT result depends on sufficient numerical convergence, particular care must be taken to ensure that the calculated properties accurately reflect real trends in materials science.~\cite{toyota2021} This is especially important for magnetic transition-metal compounds and for properties such as the MAE, which are highly sensitive to the details of the electronic structure. Accordingly, in the present work, particular care was taken to ensure numerical consistency through systematic optimization of PPs, basis-set parameters, and $k$-point sampling, as well as extensive convergence testing of MAE values (see secion~\ref{mae-sec}).

In addition to the approximations associated with the LCAO framework, DFT inherently relies on approximate XC functionals to describe electronic interactions, which directly influence binding energies and equilibrium structural parameters. For instance, it is widely recognized that commonly used functionals such as the Local Density Approximation (LDA) and the Generalized Gradient Approximation (GGA) can systematically underestimate or overestimate lattice parameters, typically within a few percent. These deviations are intrinsic to standard DFT calculations and generally do not affect the reliability of predicted structural or magnetic trends, although they should be considered when performing direct comparisons with experimental data. Furthermore, it should also be emphasized that quantitative differences between results obtained using different DFT codes are unavoidable, even when identical XC functionals are adopted. These differences originate from intrinsic methodological aspects, such as the choice of basis sets, PPs, real-space or reciprocal-space representations, and numerical implementations of the same XC functionals. Although these factors may produce moderate variations in absolute values, extensive benchmarking studies have shown that robust physical trends are typically reproduced across different computational approaches. 
For example, several benchmark studies have demonstrated that LCAO DFT implementations can produce band structures, MMs, and MAEs in close agreement with plane-wave methods when adequately converged~\cite{dftusoc,validation,u3o8,laurent}. For example, Smidstrup et al. showed comparable MAE results obtained with LCAO-based and plane-wave methods for various magnetic systems, illustrating that real-space localized basis methods can achieve quantitative accuracy for anisotropy calculations when numerical parameters are carefully controlled~\cite{quantumatk}. Additionally, technical documentation of LCAO codes (e.g., QuantumATK and SIESTA) discusses systematic convergence of MAE with respect to $k$-point sampling and basis-set quality, reinforcing that such methods can converge relevant quantities to high precision. Consequently, minor numerical discrepancies between codes are not expected to affect the reliability of the composition-dependent trends identified in the present study.
%New for the reply

\section{Results and discussion}\label{results-sec}

\subsection{DFT structural optimizations}\label{geom-sec}

As a first step, we performed ionic optimization of the fully ordered Mn$_2$RuGa alloy~($x_p$=0). It is important to note that this alloy displays distinct spin alignments across its two Mn sublattices, which makes a careful initialization of the density matrix essential. This magnetic initialization was consistently applied to all geometric configurations. After relaxation, we obtained an in--plane lattice constants of $a_0 = b_0 = 6.00$~\AA, in good agreement with the experimental value of 5.97~\AA~\cite{exp-mrg-latt}. The optimized out--of--plane parameter $c_0 = c_s/3$ was also found to be 6.00~\AA, where the division by 3 reflects the size of the simulation supercell ---unless otherwise specified, all references to $c$ throughout this work correspond to this normalized value. Te value of $c_0$ confirms the absence of tetragonal distortion in the relaxed structure. Subsequently, the fully optimized, ordered Mn$_2$RuGa structure was used as a reference to generate the 4096 possible configurations spanning all Ru concentrations, as described in Section~\ref{theory-sec}. For each of these structures, we performed a SCF single-point energy calculation. To identify and retain only the unique configurations, we applied an energy-based filtering criterion: structures were considered equivalent if their total energies differed by less than 1$\cdot$10$^{-3}$~eV. This procedure reduced the dataset to 603 inequivalent configurations, all of which were subsequently subjected to full ionic relaxation. This entire procedure is schematically illustrated in Figure~\ref{wf-fig}, which outlines the generation, filtering, and optimization workflow used to obtain the final set of inequivalent configurations. 

\begin{figure}
\centering
\epsfig{file=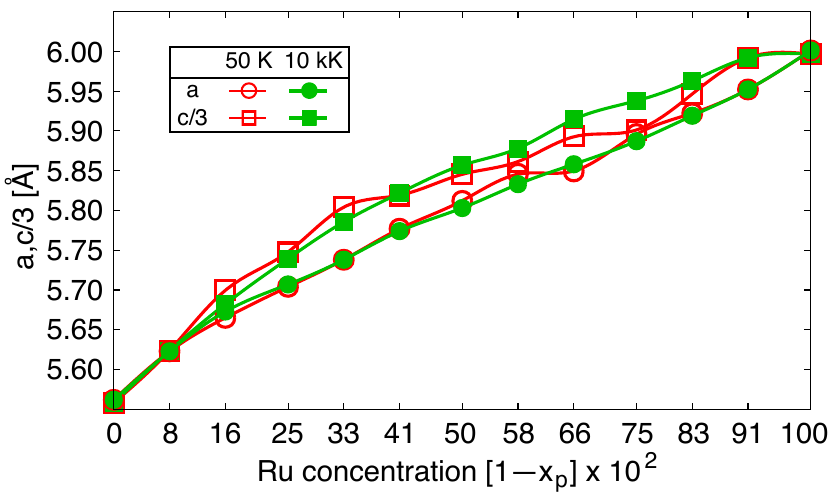,scale=0.58}
\caption{(Colour online) Schematic representation of the dependency of the in--plane and out--of--plane lattice parameters $a$ and $c$ with Ru content within the Mn$_2$Ru$_{1-x_p}$Ga supercell as a function of Ru content for two Boltzmann factors: $kT$=1~meV~(50~K) and 1~eV~(10~kK), red and green symbols, respectively. Solid lines are provided as guides to the eye.}\label{latt-fig}
\end{figure}

Figure~\ref{latt-fig} shows the dependency of the in--plane~(filled and empty circles) and out--of--plane~(filled and empty squares) lattice parameters as a function of Ru content. The first trend observed is a general increase in both lattice parameters with increasing Ru concentration, ranging from 5.58~\AA\ for Mn$_2$Ga to 6.00~\AA\ for the fully ordered Mn$_2$RuGa alloy. This behavior can be attributed to the incorporation of Ru atoms at specific lattice sites, which locally expand the Mn sublattice and lead to overall lattice enlargement inducing anisotropic changes in $a$ and $c$ lattice parameters. For low Ru content ($[1-x_p] \times 10^{2}= $ 0 to 8), the supercell retains an approximately cubic symmetry, indicating that the incorporation of up to two Ru atoms does not significantly distort the structure. However, as the Ru content increases further~($[1-x_p] \times 10^{2}= $ 16 to 91), a pronounced expansion along the out--of--plane direction is observed, while the in--plane parameter increases more moderately. In this regime, the difference between out--of--plane and in--plane lattice constants reaches a nearly constant value of $\Delta_{c-a}$$\sim$0.6~\AA\, signaling the emergence of a tetragonal distortion. From an atomic perspective, the observed preferential out-of-plane expansion of the lattice with increasing Ru content can be understood in terms of the progressive occupation of the ``vacant'' Ru sites in the Mn$_2$Ga lattice. As Ru atoms are incorporated, local structural relaxations induce an elongation along the c-axis, reflecting the interplay between atomic sizes and bonding preferences in the tetragonally distorted lattice. In particular, Ru atoms occupying asymmetric or less constrained lattice environments generate local stress that is more effectively relieved along the out-of-plane direction, where the lattice exhibits greater structural flexibility. Additionally, due to the nature of the inverse Heusler structure, the coordination environment along the z-axis may allow for more pronounced relaxation compared to the in-plane directions. This anisotropic relaxation, coupled with the configurational disorder intrinsic to metastable distributions of Ru atoms, leads to an averaged structural response that favors out-of-plane lattice expansion. This tetragonality is closely linked to the onset of MAE, as will be discussed in Section~\ref{mae-sec}, and is further supported by the PDOS analysis in Section~\ref{dos-sec}. Also shown in the figure is that the two values of $kT$ used to compute statistical averages across Ru concentrations lead to very similar trends for both the in-plane and out-of-plane lattice parameters. This indicates that, despite including configurations of higher energy when $kT$=1~eV is used, the average geometric behavior remains nearly unchanged. Such insensitivity to thermal weighting suggests a high degree of structural robustness in the Mn$_2$Ru$_{1-x_p}$Ga alloy: the lattice parameters and bonding environments are relatively unaffected by the presence of metastable configurations.

To conclude, it is important to clarify the comparison between thin-film structural data and the present bulk supercell calculations. Direct comparison of calculated lattice parameters with thin-film experiments must be performed with caution. In thin films, the in-plane lattice constants are typically constrained by the substrate, while the out-of-plane parameter can relax depending on epitaxial strain conditions. In contrast, the present calculations correspond to fully relaxed bulk configurations without substrate-induced constraints, and therefore quantitative deviations from thin-film lattice parameters are expected. Nevertheless, the trends observed in lattice distortions and MAEs reflect intrinsic bulk properties of the Mn$_2$Ru$_{1-x_p}$Ga system and remain physically meaningful and robust.

\subsection{Magnetic moments}\label{mm-sec}    

\begin{figure}
\centering
\epsfig{file=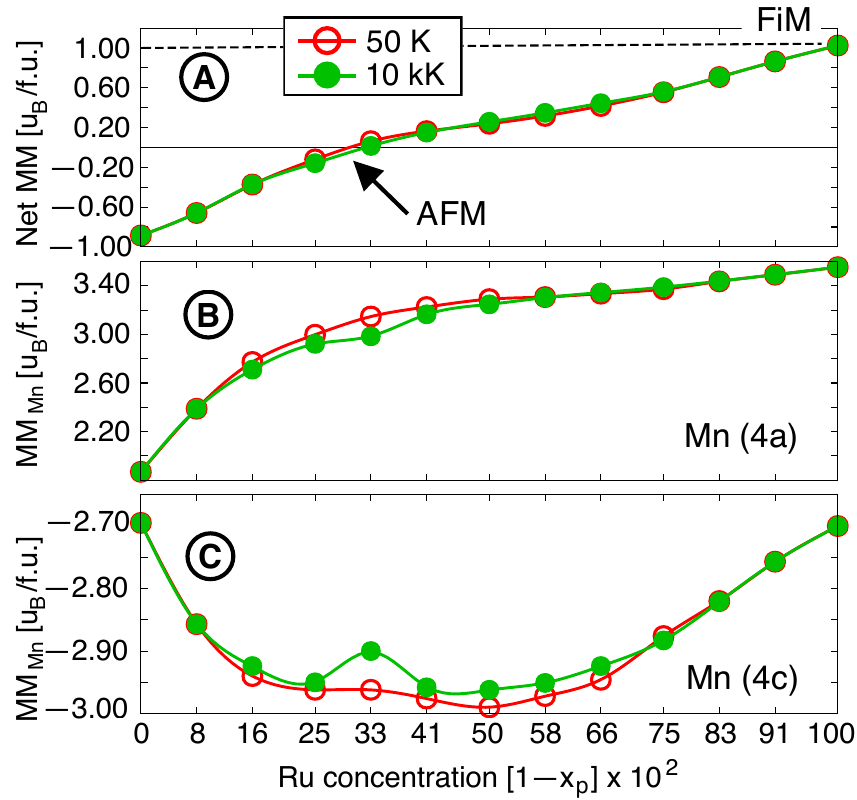,scale=0.55}
\caption{(Colour online) (A) Average net MM per formula unit of Mn$_2$Ru$_{1-x_p}$Ga as a function of Ru content for two Boltzmann factors: $kT$=1~meV~(50~K) and 1~eV~(10~kK), empty red and filled green circles, respectively. The text reports the MM values associated with the two distinct magnetic phases that emerge at both intermediate and extreme Ru concentrations: ferrimagnetic and antiferromagnetic; (B) and (C) show the average MMs of Mn atoms occupying the 4a and 4c Wyckoff positions, respectively, as a function of Ru content with the same color code as in (A). Solid lines are provided as guides to the eye.}\label{mms-fig}
\end{figure}

Figure~\ref{mms-fig}--A displays the average net MM per formula unit (MM/f.u.) of the Mn$_2$Ru$_{1-x_p}$Ga alloy as a function of Ru content, computed using Boltzmann weighting at $kT$=1~meV and 1~eV (empty red and filled green circles, respectively). According to the Slater–Pauling rule~\cite{grafa}, cubic Mn$_2$Ga and Mn$_2$RuGa should exhibit MMs of --1$\mu_B$/f.u. and +1$\mu_B$/f.u., respectively, characteristic of ideal half--metallic behavior. For $1-x_p=0$ --corresponding to cubic Mn$_2$Ga--we obtained a net moment of --0.83$\mu_B$/f.u., in good agreement with the theoretical prediction. With increasing Ru concentration, the net MM/f.u. rises monotonically, reaching +0.90$\mu_B$/f.u. for full Mn$_2$RuGa alloy. The composition at which the net MM/f.u. vanishes is approximately at $\sim 30$\%, corresponding to a configuration where both Mn sublattices  exhibit equal and opposite MMs. Thus, apart from the fully compensated composition at this specific Ru concentration, the alloy exhibits ferrimagnetic~(FiM) behavior across the entire substitution range. Two distinct magnetic regimes are identified, separated near 33\% Ru content. Below this threshold, Ru and Ga atoms contribute negatively to the net MM per formula unit (MM/f.u.), whereas above it, their contributions become positive.

Figure~\ref{mms-fig}–(B,C) shows the MM of the Mn atoms located at the 4a and 4c Wyckoff positions, respectively, as a function of Ru content in the Mn$_2$Ru$_{1-x_p}$Ga alloy. The data reveal a significant difference in sensitivity to Ru concentration between the two Mn sublattices. Specifically, the Mn(4a) sublattice exhibits a much larger MM variation, with a dispersion of 1.7~$\mu_B$, compared to only 0.3~$\mu_B$ for the Mn(4c) sublattice. This highlights that Mn atoms at 4a sites are more strongly affected by the presence and distribution of Ru atoms, suggesting that the magnetic behavior of the alloy can be effectively tuned through selective Ru substitution. The lowest MM for Mn(4a) is observed at $1-x_p$=0~(Mn$_2$Ga) while the highest corresponds to $1-x_p$=1. In contrast, the Mn(4c) sublattice shows a relatively stable MM with a subtle minimum around 50\%, at which point the MMs of both Mn sublattices approach similar values, with a difference of only $\sim$0.25~$\mu_B$. This convergence suggests a partial compensation of MMs, which plays a role in the net MM behavior described previously. It is worth noting that, similar to the lattice parameters, the MMs are only marginally affected by the Boltzmann weighting temperature. Both $kT$ produce nearly identical trends and values across all Ru concentrations, indicating that these properties are relatively insensitive to the thermal averaging scheme. 

To gain further insight into the electronic redistribution mechanisms driving the magnetic and structural evolution across the Mn$_2$Ru$_{1-x_p}$Ga series, we analyzed the Mulliken charges of each atomic species. The results reveal that Mn atoms act as electron donors throughout the entire composition range. At 0\% Ru content, the Mn atoms located at 4a and 4c Wyckoff sites exhibit charges of approximately 6.85$e$ and 6.70$e$, respectively, indicating a net loss relative to their nominal valence of 7$e$~(derived from the 4$s^2$3$d^5$ electronic configuration). As Ru content increases to 100\%, the corresponding values shift to 6.86$e$ for Mn(4a) and 6.90$e$ for Mn(4c), suggesting a subtle charge redistribution between sublattices. Ga atoms, by contrast, consistently act as electron acceptors: their charge decreases from 3.45$e$ at 0\% Ru to 3.20$e$ at full substitution, indicating increasing electron withdrawal with higher Ru content. Ru atoms remain relatively charge-neutral, with a slight tendency to act as electron acceptors~($\approx$ 8.05$e$), and exhibit only minimal variation across the compositional range. These results confirm that Mn is the primary donor species, while Ga and, to a lesser extent, Ru accommodate the transferred charge—an effect that plays a key role in stabilizing the magnetic and electronic structure of the alloy. In general, due to the alignment of the $d$-orbital energies (as will be discussed in the following section), charge transfer and redistribution occur both within and between the Mn sublattices, leading to a continuous variation in the MMs of the Mn atoms across the entire Ru concentration range.

\subsection{Projected density of states}\label{dos-sec}

\begin{figure}
\centering
\epsfig{file=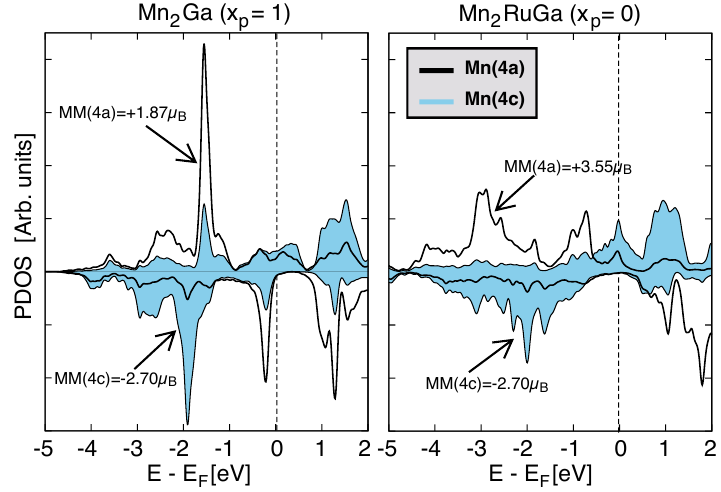,scale=0.70}
\caption{(Colour online) Spin-resolved projected density of states of Mn atoms of full Mn$_2$Ga~(left) and Mn$_2$RuGa~(right). Solid black line depicts the PDOS of Mn(4a) and filled curve those of Mn(4c). The local MMs of the Mn atoms in both alloys are indicated within the subfigures, with arrows pointing to the associated PDOS contributions of each  sublattice.}\label{100_0-fig}
\end{figure}

To assess the half-metallic character of the system, Figure~\ref{100_0-fig} displays the PDOS onto the two Mn atomic sublattices for Mn$_2$Ga~(left) and M$_2$RuGa~(right), representing the two extreme cases of Ru concentration. For Mn$_2$Ga, the Fermi level lies close to the edge of the spin-down band gap, indicating an almost half-metallic nature. In contrast, Mn$_2$RuGa shows a well-defined half-metallic behavior, with the Fermi level located deep within a spin-down gap of approximately 1~eV. In addition, we observe that the electronic structures of Mn$_2$Ga and Mn$_2$RuGa differ markedly. In the case of Mn$_2$Ga, the electronic states of the two Mn sublattices are more localized and exhibit a high degree of alignment across the energy range in both spin channels. For the spin-up states, the bands of the two inequivalent Mn atoms strongly overlap throughout the entire energy range, especially around the Fermi level and from --1 eV to --3 eV, notably at --1.6 eV. In the spin--down channel, the alignment is also observed just below the Fermi level at --0.2 eV, between --1.1 eV and --4.1 eV and notably around --2.0 eV. This orbital alignment facilitates charge transfer between Mn atoms located at different Wyckoff positions, which are separated by approximately 2.6\AA. For Mn$_2$RuGa, the electronic states of Mn(4a) and Mn(4c) posses an elevated grade of coupling at the Fermi level for the spin-up states. For these up states, the Mn(4a) present two well defined enegy range coupled with those Mn(4c), namely, from Fermi level up to --1.5 eV and from --1.5 eV up to --4.5 eV. The spin-down states of both Mn atoms overlap from around --0.6 eV up to --4.5 eV. It is worth noting that, upon examining the PDOS curves, the emergence of the spin-down gap appears to originate from charge transfer processes occurring both between the two Mn sublattices and between Mn and Ga atoms as specified within the last paragraph of section~\ref{mm-sec}. This interaction induces a Fermi level shift of approximately 1 eV from Mn$_2$Ga to Mn$_2$RuGa, accompanied by a broadening and redistribution of the spin-down electronic states. These changes ultimately stabilize the half-metallic character observed in Mn$_2$RuGa.

\begin{figure*}
\centering
\epsfig{file=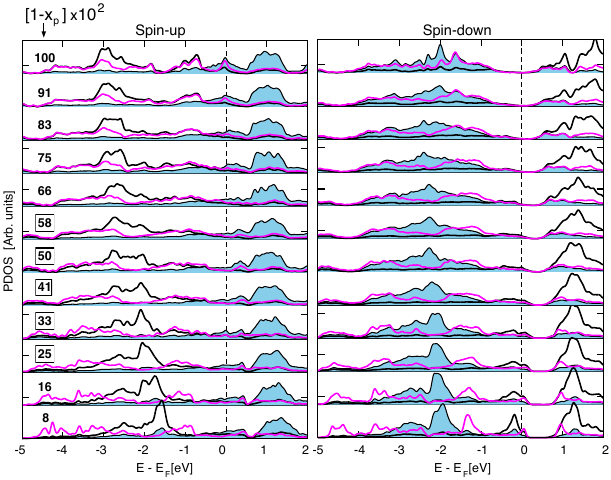,scale=1.5}
\caption{(Colour online) Average density of states for spin-up~(left) and spin-down~(right) channels projected onto Mn and Ru atoms. The solid black lines correspond to Mn(4a), the filled blue areas represent Mn(4c), and the solid pink lines correspond to Ru. The Ru concentration ([$1-x_p$]$\times$10$^{2}$) increases progressively from 8\% at the bottom to 100\% at the top of the figure. The black boxes highlight the Ru concentrations where the easy axis is out-of-plane(see Figure~\ref{mae-fig}).}\label{mn4ac+ru-fig}
\end{figure*}

Figure~\ref{mn4ac+ru-fig} displays the PDOS onto the two inequivalent Mn Wyckoff positions and onto the Ru atoms of Mn$_2$Ru$_{1-x_p}$Ga, obtained by ensemble averaging at each Ru concentration using a Boltzmann factor with $kT=1$~meV. 
At low Ru concentration~(8\%), the spin-up PDOS of Mn(4a)~(solid black line, left panel) closely resembles the spin-down PDOS of Mn(4c)~(filled blue curve, right panel), showing similar band shapes. As the Ru content increases~(up to approximately 58\%), this resemblance persists, although the electronic states become progressively more delocalized and less sharply defined. Beyond this concentration, and up to full Ru content~(100\%), the Mn(4a) spin-up PDOS~(left panel) develops two distinct subbands centered around –1~eV and –3~eV, while the Mn(4c) spin-down PDOS~(right panel, filled blue) becomes concentrated between –1~eV and –4~eV. This evolution reflects a growing electronic perturbation induced by Ru, which affects both Mn types and spin channels. A comparable pattern is observed when comparing the spin-up PDOS of Mn(4c)~(left panel, filled blue) with the spin-down PDOS of Mn(4a)~(right panel, solid black). At low Ru content, the profiles remain fairly similar, except for a prominent peak just below the Fermi level at –0.2~eV in the spin-down PDOS of Mn(4a). However, as the Ru concentration increases, a spin-up peak emerges at the Fermi level for Mn(4c), while the corresponding spin-down feature in Mn(4a) gradually diminishes. These observations support the conclusion made in the previous section: as Ru content increases, the hybridization between Mn(4a) and Mn(4c) states weakens. This suggests that Ru atoms progressively screen the interaction between the two Mn sublattices, reducing electronic overlap and magnetic coupling.

Regarding the PDOS of Ru atoms~(solid pink lines), it is evident that the Ru $d$-states, in conjunction with those of Mn, play a significant role in shaping the electronic structure near the Fermi level. As the Ru concentration increases from 8\% to 100\%, the spin-up states contribute to an enhanced density at the Fermi energy. Simultaneously, the spin-down states help maintain the gap characteristic of half-metallicity: a small peak initially present at the Fermi level progressively diminishes with increasing Ru content and eventually disappears entirely at 100\% Ru. This behavior indicates that Ru atoms are not merely passive dopants but actively participate in the formation and stabilization of the system’s half-metallic character. In addition, the Ru spin-up PDOS exhibits two distinct subbands for lower Ru content: one spanning from 0 to –2~eV, and a second from –2~eV to –4.5~eV. The first subband, extending from the Fermi level down to approximately –2~eV, displays a nearly flat profile in the region between 0 and –0.8~eV, indicating more dispersive electronic states. From –0.8~eV to –2~eV, however, the PDOS becomes significantly more pronounced, suggesting a higher density of narrower, more localized states. The second subband, from –2~eV to –4.5~eV, remains relatively constant in intensity across the entire energy range. In contrast, the Ru spin-down PDOS follows a slightly different trend: it features a more localized peak centered around –1.3~eV and a broader subband extending from –2.2~eV to –4~eV. As the Ru concentration increases, the PDOS profiles in both spin channels progressively flatten, merging the previously discussed subbands into a broader, more uniform distribution that extends from the Fermi level down to approximately –4~eV. Notably, for the spin-down channel, the prominent peak observed at lower Ru concentrations vanishes entirely—mirroring the behavior seen for the Mn species. Finally, the overlap between Ru and Mn states becomes evident within specific energy regions. For the spin-up Ru $d$-states, an initial misalignment with the Mn(4a) and Mn(4c) states is observed at low Ru concentrations. However, beginning around 50\%, these states broaden and progressively align across the full energy range. A comparable, though less pronounced, trend is observed for the spin-down channel: a modest overlap appears at low Ru content and gradually increases as the Ru concentration rises.

In general, the observed shift of the PDOS peaks toward more negative energies indicates a stabilization of the occupied states, typically arising from charge transfer, enhanced orbital hybridization, or modifications of the local chemical potential induced by variations in Ru content. Such shifts demonstrate how compositional tuning reorganizes the electronic structure, which in turn governs the redistribution of MMs and the robustness of half-metallicity.

\subsection{Magnetic anisotropy energy}\label{mae-sec}
\begin{figure}
\centering
\epsfig{file=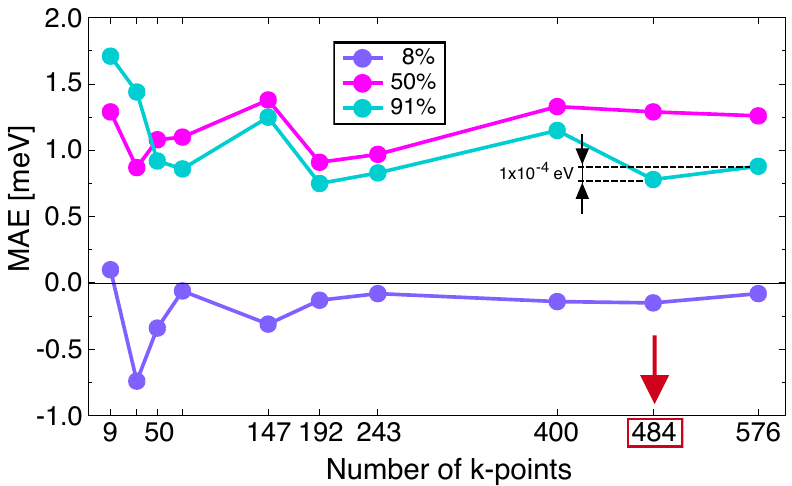,scale=0.62}
\caption{(Colour online) Magnetic anisotropy energy convergence as a function of the number of $k$-points for three representative Mn$_2$Ru$_{1-x_p}$Ga compositions. The interval between the dashed lines indicates the MAE dispersion at high $k$-point densities. The red square and arrow mark the selected $k$-point mesh used in the production calculations to ensure converged MAE values.}\label{mae-conv-fig}
\end{figure}

% New in the reply
As pointed out in Section~\ref{theory-sec}, the MAE values are of the order of a few meV. Therefore, the calculation of self-consistent total energies requires very high numerical accuracy. To ensure reliable MAE values, we performed systematic convergence tests with respect to the $k$-point sampling. For each Ru concentration, a configuration was selected to evaluate this convergence, showing similar convergence trends across all concentrations. Figure~\ref{mae-conv-fig} shows the MAE convergence for three Ru concentrations, namely 8\%, 50\%, and 91\%, corresponding to low, intermediate, and high Ru content in the Mn$_2$Ru$_{1-x_p}$Ga alloy, respectively, as a function of the number of $k$-points. The dashed black lines indicate the target MAE accuracy of $1\times10^{-4}$ eV. We observe that this tolerance is achieved for all three compositions at 484 $k$-points. Based on these results, this $k$-point mesh was adopted for all calculations reported in this work, including structural relaxations, MMs, MAE, and electronic structure analyses.
% New in the reply

% New in the reply
% New in the reply

\begin{figure*}
\centering
\epsfig{file=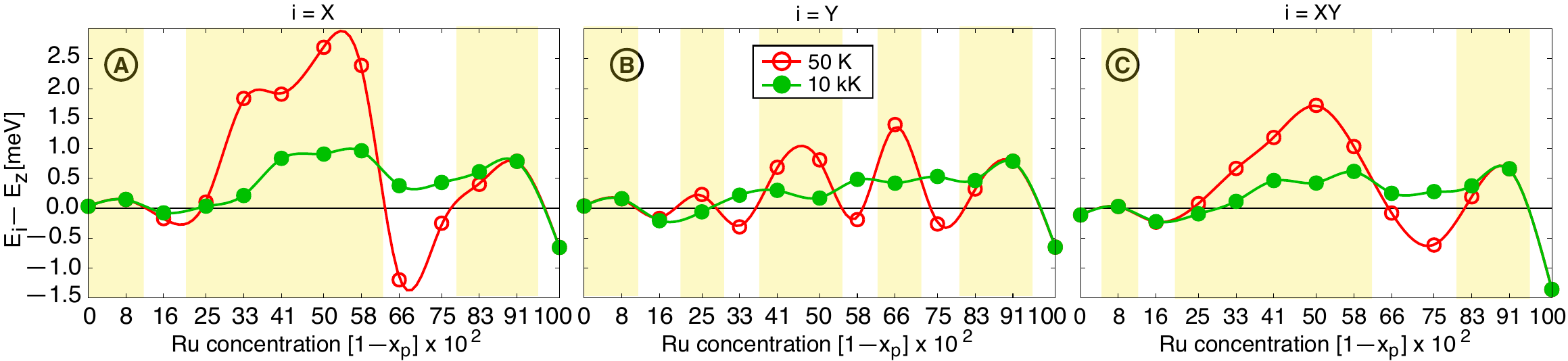,scale=0.45}
\caption{(Colour online) Average total SCF energy difference as a function of Ru content for two Boltzmann factors: $kT$=1~meV~(50~K) and 1~eV~(10~kK)~(empty red circles and filled green circles), $\Delta E_x=E_x-E_z$ (A), $\Delta E_y=E_y-E_z$ (B), $\Delta E_{xy}=E_{xy}-E_z$ (C). The light yellow-shaded regions denote out-of-plane easy axis orientations for $kT$=1~meV, while smooth solid lines serve as guides to the eye.}\label{mae-fig}
\end{figure*}

We calculated the total SCF energy of all inequivalent structures with the magnetization oriented along the cartesian directions~(X, Y, and Z), as well as the in-plane XY direction. Figure~\ref{mae-fig} shows the SCF energy differences defined as $\Delta E_i=E_i-E_z$ with $i=x,$ $y$, $xy$, plotted as a function of Ru concentration in panels A, B, and C, respectively.

In contrast to the MMs, lattice parameters, and projected density of states, $\Delta E_i$ exhibits a strong dependence on the Boltzmann weighting scheme. Specifically, using $kT$=1~meV or 1~eV to compute the weighted averages yields markedly different behaviors: at low $kT$, the energy differences display pronounced oscillations, which become smoothed out at higher $kT$. These oscillations reflect changes in the preferred magnetization direction, with the magnetic easy axis alternating between out-of-plane and in-plane orientations depending on the specific atomic configuration~(see light yellow shaded regions in the figure~\ref{mae-fig}). This behavior highlights the sensitivity of the MAE to local atomic arrangements and thermal weighting, revealing a complex interplay between structural disorder and MAE in the Mn$_2$Ru$_{1-x_p}$Ga system. It is worth noting that these oscillations vanish at $kT$=1~eV, resulting in a well-defined out-of-plane easy axis for the MAE across the entire Ru composition range—except at low Ru concentrations~(specifically 16\% and 100\%), where the easy axis remains in-plane.

Focusing first on the $kT$=1~meV case within the yellow-shaded Ru concentration regions, we observe that the largest energy differences occur for $\Delta E_x$ at 50\% and 58\%, reaching a maximum of approximately 2.4-2.7~meV~(empty red circles in Figure~\ref{mae-fig}–A). This indicates a clear preference for an out-of-plane easy axis in this range and a strong MAE compared to the nearly zero values or in-plane easy axis found in the pure Mn$2$Ga and Mn$_2$RuGa alloys. For the other SCF energy differences, the highest values are observed at 66\% for $\Delta E_y$ and at 50\% for $\Delta E_{xy}$, with values of 1.4~meV and 1.7~meV, respectively. Importantly, across the entire Ru composition range, the out-of-plane orientation is favored over the in-plane one for all $\Delta E_i$ cases—approximately 70\% of compositions for $i=x$, and 62\% for $i=y$ and $i=xy$. Moreover, the positive values of $\Delta E_i$ tend to be larger in magnitude than the negative ones, depending on the specific Ru region. For instance, in the case of $\Delta E_x$, the energy differences between the out-of-plane configuration at 58\% and the in-plane configuration at 66\% is approximately 3.5~meV. This significant energy barrier suggests that the material at these concentrations is sufficiently robust to retain spin orientation, which is desirable for memory storage applications. A similar trend is seen for $\Delta E_{xy}$, where the energy difference between the out-of-plane configuration at 50\% and the in-plane configuration at 75\% reaches 2.3~meV. In contrast, for $\Delta E_y$, the transition between out-of-plane and in-plane easy axes is smoother, with smaller energy barriers due to larger oscillations in MAE across the composition range. When considering $kT$=1~eV, although the qualitative transition behavior between easy axis orientations remains consistent with the lower thermal factor case, the overall MAE values are substantially reduced—reaching at most 1~meV for $\Delta E_x$ and dropping to as low as 0.5~meV for both $\Delta E_y$ and $\Delta E_{xy}$.  
          
\section{Exploratory Analysis of Structural and Electronic Descriptors}\label{pca-sec}          

\begin{figure}
\centering
\epsfig{file=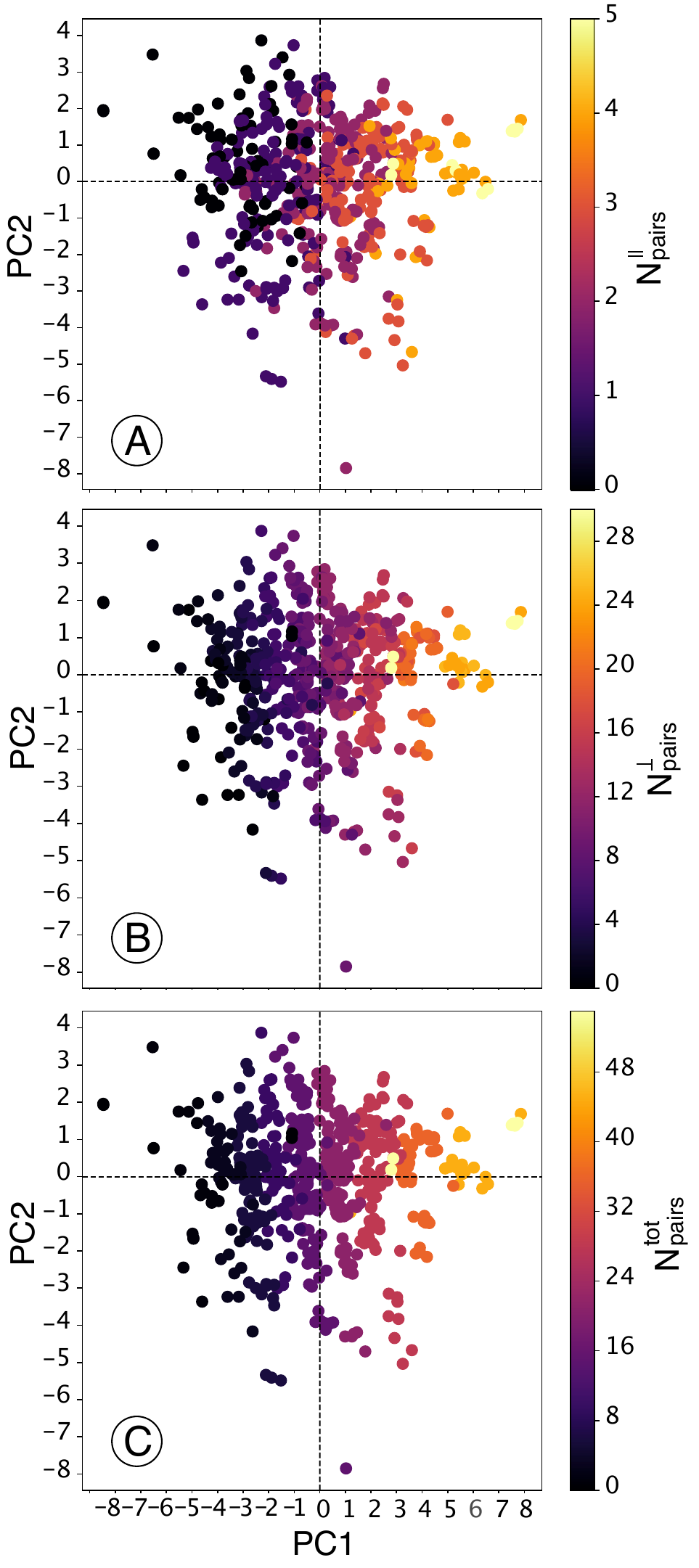,scale=0.43}
\caption{(Colour online) (A) Projection of all configurations onto the first two principal components (PC1 on the x-axis, PC2 on the y-axis), with the color indicating the number of in-plane vacancy pairs ($N_{\rm pairs}^\parallel$); (B) Same projection, colored by the number of out-of-plane vacancy pairs ($N_{\rm pairs}^\perp$); (C) Same projection, colored by the total number of vacancy pairs ($N_{\rm pairs}^{\rm tot}$). Each point represents a distinct geometric configuration.}\label{npairs-fig}
\end{figure}

\begin{figure*}
\centering
\epsfig{file=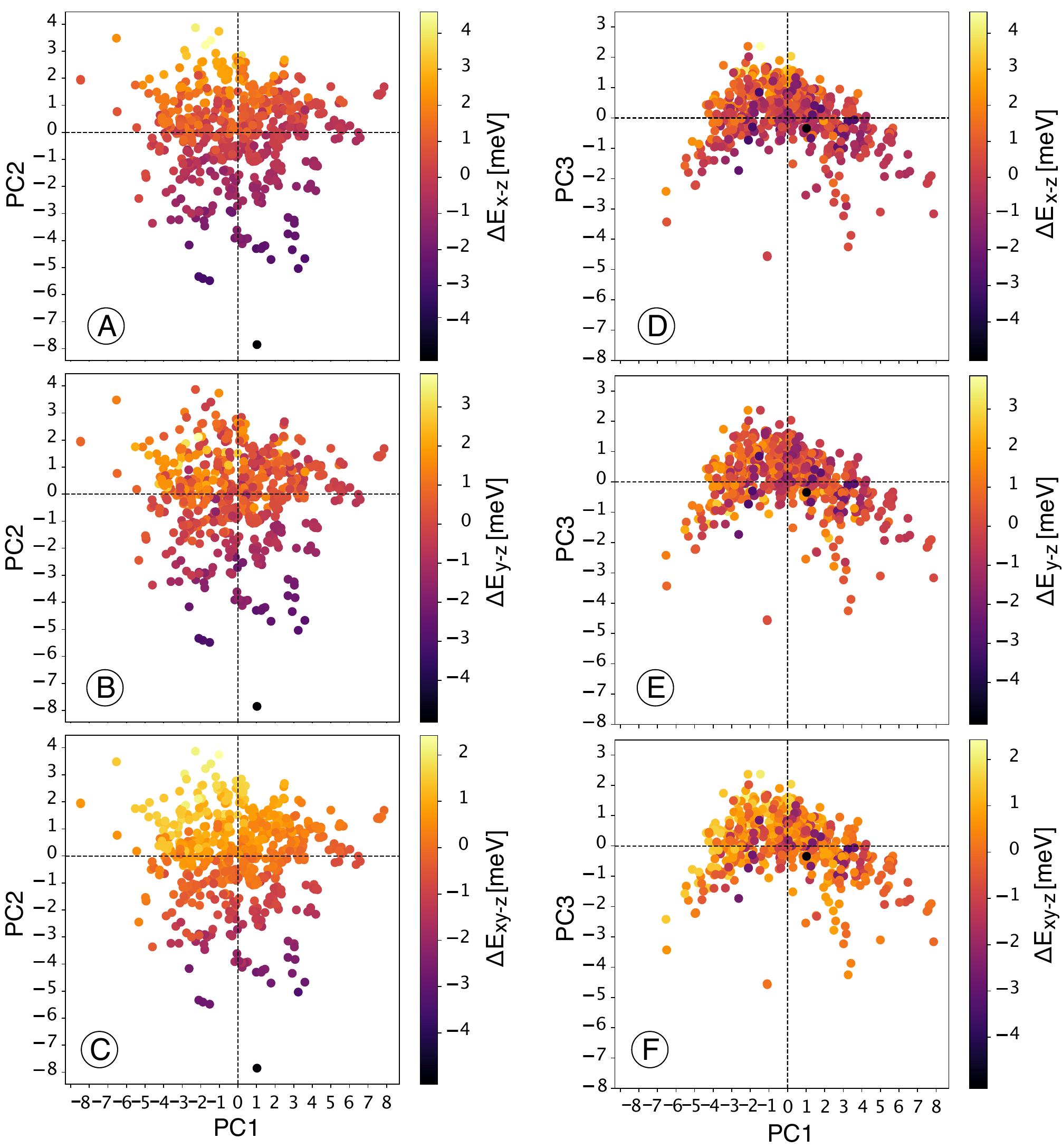,scale=0.43}
\caption{(Colour online) Projection of the dataset onto the first three principal components. Panels (A–C) and (D–F) display the PC1–PC2 and PC1–PC3 projections, respectively. The color scale represents the energy differences $\Delta E_{x-z}$, $\Delta E_{y-z}$, and $\Delta E_{xy-z}$. Each point corresponds to a distinct geometric configuration.}\label{scatter_mae-fig}
\end{figure*}

In this section, we employ unsupervised multivariate analysis via PCA to investigate the interconnections between Ru vacancy pairs, mean pairwise distance between vacancies, and total SCF energy difference for x, y, z and xy spin quantization axes. For this purpose, nineteen descriptors were selected, as detailed in section~\ref{theory-sec}, which are sufficient to capture the qualitative trends of the magnetic behavior of the Mn$_2$Ru$_{1-x_p}$Ga alloy. The resulting loadings heat map (figure~\ref{heat-map-fig}) shows that PC1 is primarily influenced by the number of vacancy pairs in-plane, out-of-plane, total pairs, and three-dimensional pairs, as well as by the total and planar radii of gyration. Secondary contributions arise from the total SCF energy differences ($\Delta E_{x-z}$, $\Delta E_{y-z}$, $\Delta E_{xy-z}$ and Mn(4c) MMs), while tertiary contributions are mainly associated with the mean pairwise distances between vacancies along the in-plane and out-of-plane directions. Additional PCs contributions of lesser magnitude are indicated in the figure. PC1 accounts for 36\% of the total variance, identifying these descriptors as the main drivers of variability in the dataset. PC2, which explains 14\% of the variance, is mainly shaped by the same energy differences, thereby complementing and reinforcing the correlations highlighted in PC1. As we have pointed out in section~\ref{theory-sec}, loadings act as coefficients that quantify the contribution of each descriptor to the PCs, such that descriptors with larger absolute loadings exert a stronger influence on the position of each configuration in the reduced PC space. This weighting mechanism enables the identification of geometric and electronic properties that govern the emergent behavior of the alloy. Accordingly, the 603 points plotted in the two-dimensional projections of figures~\ref{npairs-fig} and \ref{scatter_mae-fig} represent individual geometric configurations positioned according to their coordinates along PC1 and PC2/PC3. Configurations shifted towards positive values along these axes reflect stronger contributions from $N_{\rm pairs}^{\parallel (\perp)}$, $R_{\rm g}^{\parallel (tot)}$, SCF energy differences, and in-plane and out-of-plane $\langle d^{\rm PBC} \rangle$. The accompanying color bars in each panel further disentangle these effects by explicitly encoding the magnitude of the property under consideration, allowing a direct visual interpretation of how descriptors shape the overall magnetic and structural trends.      

In Figure~\ref{npairs-fig}(A–C), the color bars indicate the number of in-plane ($N_{\rm pairs}^\parallel$), out-of-plane ($N_{\rm pairs}^\perp$), and total vacancy pairs ($N_{\rm pairs}^{\rm tot}$), respectively. Configurations located farther from the origin along PC1 and PC2 are more strongly influenced by descriptors with larger loadings. In particular, configurations located at top-right quadrant of these plots predominantly exhibit out-of-plane MAEs and correspond to Ru contents between 25\% and 58\%. Focusing on this region, we observe a clear relationship between the MAE, Ru composition, and the spatial arrangement of vacancy pairs. In Figure~\ref{npairs-fig}(A), configurations with in-plane vacancy pair numbers predominantly between three and five (orange to yellow in the color scale) are located at positive PC1 and PC2 values. This indicates that, for this range of Ru contents, structures exhibiting an out-of-plane easy axis typically contain approximately this number of in-plane vacancy pairs, suggesting a moderate degree of spatial correlation among the vacancies rather than a completely dispersed distribution. In contrast, structures with negative PC1 or PC2 values are only weakly influenced by these descriptors. In this region, the small number of vacancy pairs indicates that the vacancies remain largely isolated, leading to weaker modifications of the local electronic environment. As a consequence, the Mn(4c) MMs show little correlation with the MAE, and only minor variations in the total SCF energy are observed when changing the quantization axis. A similar behavior is observed in panel (B), where the number of out-of-plane vacancy pairs spans a wider range (12–28 pairs, from dark orange to yellow). These larger values indicate a higher degree of spatial correlation among vacancies along the out-of-plane direction, meaning that vacancies tend to occur closer to each other rather than remaining isolated. Such configurations produce a stronger perturbation of the local crystal symmetry along the z-axis. As a consequence, the electronic environment becomes increasingly anisotropic, enhancing the sensitivity of the SOC to the magnetization direction through changes in the orbital hybridization of neighboring atoms. In Figure~\ref{npairs-fig}(C), the color scale represents the total number of vacancy pairs per configuration, which now spans a broader range (16–40 pairs). These larger values indicate an overall increase in the spatial correlation of vacancies within the lattice. In this representation, configurations located in the upper-right quadrant combine both in-plane and out-of-plane vacancy correlations, suggesting that the MAE is influenced by the joint contribution of both. Rather than being controlled by a single directional descriptor, the MAE appears to reflect the global topology of the vacancy network. In particular, configurations with higher total numbers of vacancy pairs tend to correspond to stronger out-of-plane MAEs, highlighting the role of vacancy clustering in modulating the anisotropy. It is worth noting that the simulation cell is elongated along the z-direction, which increases the periodic separation between vacancies along this axis. Therefore, the occurrence of significant numbers of out-of-plane vacancy pairs reflects genuine spatial correlations rather than a trivial geometrical effect of the supercell.

To further assess the influence of the descriptors contributing to PC1, PC2, and PC3 (first three columns in the heatmap shown in Fig.~\ref{heat-map-fig}), we plot in Fig.~\ref{scatter_mae-fig}(A–F) the PC1–PC2 and PC1–PC3 scatter projections. The PC1–PC2 panels display a point distribution similar to that shown in Fig.~\ref{npairs-fig}. The color scale of all panels represents the SCF total energy differences calculated along the high-symmetry magnetization directions x, y, z, and xy. As usual, positive energy differences indicate a preference for an out-of-plane easy axis, whereas negative values correspond to an in-plane orientation. As discussed in Sec.~\ref{mae-sec}, the largest $\Delta E$ values among all configurations are obtained for the $E_x - E_z$ energy difference, which is primarily associated with configurations located at PC2 $>$ 0 in panels (A–C). Consistently with Fig.~\ref{npairs-fig}, the first quadrant—mainly corresponding to configurations with Ru contents between 25\% and 58\%—is dominated by positive energy differences, indicating a clear tendency toward an out-of-plane easy axis. This behavior can be rationalized in terms of the spatial organization of the Ru vacancies encoded in the descriptors contributing to PC1. In this region, the lattice contains a relatively large number of vacancies whose spatial correlations are reflected in the descriptors $N_{\rm pairs}$ and $R_g$, forming correlated pairs both in-plane and out-of-plane, which leads to an extended vacancy network within the lattice. Such arrangements introduce stronger local symmetry breaking and modify the hybridization of neighboring atoms' orbitals, which enhances the sensitivity of the electronic structure to the magnetization direction and ultimately stabilizes an out-of-plane MAE, consistent with the positive energy differences observed for PC2 $>$ 0. By contrast, configurations with negative PC1 and/or PC2 values typically correspond either to low vacancy concentrations or to more isolated vacancy distributions. In these cases, the reduced number of vacancy pairs and the smaller spatial extent of the vacancy clusters (smaller $R_g$) produce weaker perturbations of the local electronic environment, resulting in smaller anisotropy energies and a weaker tendency toward an out-of-plane easy axis.
 
In Fig.~\ref{scatter_mae-fig}(D–F), the PC1–PC3 scatter projections highlight the interplay between the total number and spatial arrangement of vacancies~(encoded in PC1 via $N_{\rm pairs}$ and $R_g$) and the mean in-plane and out-of-plane vacancy separations, $\langle d^{PBC}_{\rm in-plane}\rangle$ and $\langle d^{PBC}_{\rm out-of-plane} \rangle$, as well as slight contributions from the tetragonal distortion $\Delta_{c-a}$ captured in PC3. Configurations located at PC3 $>$ 0 are associated with larger values of $\langle d^{PBC}_{\rm out-of-plane} \rangle$ and moderately large in-plane separations, indicating that the vacancies are not only numerous but also more spatially extended along the z-axis, forming a three-dimensional correlated network. This extended distribution further enhances the breaking of the local symmetry along the out-of-plane direction, increasing the sensitivity of the SOC to the underlying lattice geometry. Consequently, configurations with PC3 $>$ 0 predominantly exhibit an out-of-plane easy axis, consistent with the positive SCF energy differences observed in these panels. In contrast, configurations with negative PC3 values tend to have smaller inter-vacancy distances or less correlated vacancy arrangements, resulting in weaker symmetry breaking and a reduced tendency for out-of-plane MAE.

In summary, the analysis shows that the MAE of Mn$_2$Ru$_{1-x_p}$Ga is primarily dictated by the spatial correlations of Ru vacancies rather than by the nominal Ru concentration alone. Configurations with moderate Ru content~(around 25–58\%) contain a significant number of correlated vacancy pairs, both in-plane and out-of-plane, which can further aggregate into vacancy clusters forming extended three-dimensional networks. These spatially correlated arrangements break the local lattice symmetry along the out-of-plane direction, altering the crystal-field splitting and orbital hybridization of neighboring atoms, which increases the sensitivity of the SOC to the magnetization direction. As a result, configurations with well-formed vacancy pairs and clusters predominantly exhibit an out-of-plane easy axis, as reflected in the positive SCF energy differences. Conversely, configurations with fewer, isolated, or less correlated vacancies display smaller inter-vacancy distances and weaker spatial correlations, producing only minor symmetry breaking and favoring an in-plane easy axis. The combination of the total number of vacancy pairs ($N_{\rm pairs}$), their spatial correlations~($R_g$), the in-plane and out-of-plane separations~($\langle d^{PBC}_{\rm in-plane} \rangle$, $\langle d^{PBC}_{\rm out-of-plane} \rangle$), and minor tetragonal distortions~($\Delta_{c-a}$) provides a consistent physical framework for understanding how the topology and clustering of vacancies control the MAE in this system.
    
\section{Conclusions}\label{conclusions-sec}

In this work, we have carried out a systematic first-principles investigation of the magnetic and electronic properties of the full Heusler Mn$_2$Ru$_{1-x_p}$Ga alloy, focusing on the role of Ru concentration ($x_p = 0.0834 \cdot p$, $p = 0, \dots, 12$) in tuning key features such as lattice distortion, magnetic ordering, half-metallicity, and MAE. Our methodology combined density functional theory within a high-throughput framework with data-driven techniques from artificial intelligence and multivariate analysis. Over 4000 atomic configurations were initially generated for the 13 compositions under study and filtered by total energy single-point calculations, ultimately yielding 603 inequivalent metastable configurations that were fully relaxed. To incorporate the effects of configurational disorder and thermal fluctuations in a statistical manner, Boltzmann-weighted averages of all properties were computed at two effective thermal energy values, $kT$ = 1 meV and $kT$ = 1 eV. This integrated strategy allowed us to systematically capture the sensitivity of the system to local atomic arrangements and assess their impact on the magnetic responses.

Structurally, the relaxation process revealed a consistent anisotropic lattice expansion across the Ru composition range, with a pronounced elongation of the out-of-plane lattice parameter~($c$) relative to the in-plane parameters~($a$). This distortion leads to a symmetry breaking from cubic to tetragonal geometry, which is known to enhance perpendicular MAE. Our analysis of atomic MMs showed that the ferrimagnetic coupling between Mn atoms at 4a and 4c Wyckoff positions becomes nearly compensated below 50\% Ru content, leading to a net MM close to zero around 30\%—a feature of interest for AFM spintronics. Moreover, the electronic density of states revealed that the half-metallic character of the alloy is selectively preserved at both low and high Ru content, whereas intermediate concentrations disrupt the spin polarization due to electronic state broadening and orbital hybridization. From the PDOS, we observed a clear redistribution and delocalization of $d$-orbital states between Mn and Ru atoms with increasing Ru content, resulting in broader and more hybridized bands that directly impact MAE and spin polarization. Regarding the MAE, the analysis of the SCF energy differences among spin orientations ($E_i - E_z$ with $i = x$, $y$, $xy$) shows that the emergence of an out-of-plane easy axis is most pronounced at intermediate Ru concentrations (25–58\%). This behavior can be rationalized by the spatial organization of Ru vacancies: at these concentrations, a significant number of correlated vacancy pairs and extended vacancy clusters form both in-plane and out-of-plane, breaking the local lattice symmetry along the out-of-plane direction. The resulting enhancement of orbital anisotropy and spin–orbit coupling stabilizes the out-of-plane easy axis. At lower or higher Ru contents, vacancies are fewer or more isolated, producing weaker symmetry breaking and favoring in-plane anisotropy or nearly isotropic behavior. These trends persist across both low- and high-temperature Boltzmann distributions, although the absolute MAE values decrease at higher $kT$, as expected.

In summary, our findings demonstrate that by carefully tuning the Ru concentration in Mn$_2$Ru$_{1-x_p}$Ga, it is possible to engineer a material with controllable MAE, tunable half-metallicity, and nearly compensated ferrimagnetism. These combined features make this alloy a strong candidate for spintronic applications, particularly in magnetic tunnel junctions, MRAM devices, and high-density perpendicular magnetic storage technologies.

\section{Acknowledgments}
The author is grateful to Dr. Unai Atxitia for helpful discussions. We acknowledge the Severo Ochoa Center of Excellence Program through Grant CEX2024-001445-S. R. C. also acknowledges the computer resources at MareNostrum5 GPP and the technical support provided by the Barcelona Supercomputing Centre through Red Española de Supercomputación~(Grant No. QHS-2024-3-0011).

\bibliographystyle{unsrt}

\end{document}